\documentclass[aps,prc,reprint,groupedaddress]{revtex4-1}
\usepackage{graphicx} 
\usepackage{dcolumn} 
\usepackage{bm}
\usepackage{hyperref}

\begin{document}

\title{Preequilibrium particle emissions and in-medium effects on the pion production in heavy-ion collisions}
\author{Zhao-Qing Feng}
\email{fengzhq@impcas.ac.cn}

\affiliation{Institute of Modern Physics, Chinese Academy of Sciences, Lanzhou 730000, People's Republic of China }

\date{\today}
\begin{abstract}
Within the framework of the Lanzhou quantum molecular dynamics (LQMD) transport model, pion dynamics in heavy-ion collisions near threshold energies and the emission of preequilibrium particles (nucleons and light complex fragments) have been investigated. A density, momentum and isospin dependent pion-nucleon potential based on the $\Delta$-hole model is implemented in the transport approach, which slightly leads to the increase of the $\pi^{-}/\pi^{+}$ ratio, but reduces the total pion yields. It is found that a bump structure of the $\pi^{-}/\pi^{+}$ ratio in the kinetic energy spectra appears at the pion energy close to the $\Delta$(1232) resonance region. The yield ratios of neutrons to protons from the squeeze-out particles perpendicular to the reaction plane are sensitive to the stiffness of nuclear symmetry energy, in particular at the high-momentum (kinetic energy) tails.

\begin{description}
\item[PACS number(s)]
21.65.Ef, 21.65.Jk, 24.10.Jv
\end{description}
\end{abstract}

\maketitle

\section{Introduction}

The equation of state (EOS) in dense nuclear matter is a basic quantity in understanding the structure of neutron star, the cooling process of protoneutron stars, the nucleosynthesis during supernova explosions of massive stars etc \cite{St05}. The EOS at zero temperature is usually expressed via the energy per nucleon in the isospin asymmetric nuclear matter as $E(\rho,\delta)=E(\rho,\delta=0)+E_{\textrm{sym}}(\rho)\delta^{2}+\textsc{O}(\delta^{2})$ in terms of baryon density $\rho=\rho_{n}+\rho_{p}$, relative neutron excess $\delta=(\rho_{n}-\rho_{p})/(\rho_{n}+\rho_{p})$, energy per nucleon in a symmetric nuclear matter $E(\rho,\delta=0)$ and nuclear symmetry energy $E_{\textrm{sym}}=\frac{1}{2}\frac{\partial^{2}E(\rho,\delta)}{\partial\delta^{2}}\mid_{\delta=0}$. From that, one can easily get the physical quantities of nuclear matter, such as pressure, chemical potential, imcompressibility etc. Heavy-ion collisions with neutron-rich beams at intermediate energies are a unique tool to extract the information of isospin asymmetric nuclear matter under extreme conditions in terrestrial laboratories, such as high baryon density, high temperature, large isospin asymmetry etc.

The density dependence of the symmetry energy around and below the normal densities has been roughly constrained from heavy-ion collisions \cite{Ba05,Li08}. However, the high density information on the symmetry energy is poorly known up to now. Besides nucleonic observables, particles produced in heavy-ion collisions would be preferable probes for extracting the information of high-density phase diagram. The ratios of isospin particles produced in heavy-ion collisions such as $\pi^{-}/\pi^{+}$, $K^{0}/K^{+}$, $\Sigma^{-}/\Sigma^{+}$ etc \cite{Li02,Li05,Fe05,To10,Pr10,Fe13}, neutral particles such hard photons \cite{Yo08}, $\eta$ etc \cite{Yo13}, and the flow difference of isospin particles \cite{Gi10,Fe12a} have been proposed as sensitive probes for extracting the high-density behavior of the nuclear symmetry energy. Interplay of the mean-field potentials and corrections on threshold energies associated with production cross sections of particles impacts the particle dynamics and isospin ratios. The in-medium effects on pions and $\Delta(1232)$ dynamics in heavy-ion collisions have been studied with transport models \cite{Xi93,Ho14,So15}. Dynamics of particles produced in heavy-ion collisions energies is complicated, which is related to the production and transportation in dense nuclear medium.

In this work, the isospin dynamics of pions and preequilibrium particles in heavy-ion collisions is to be investigated with the Lanzhou quantum molecular dynamics (LQMD) transport model. The article is organized as follows. In section II we give a brief description of the LQMD model. Calculated results and discussions are presented in section III. Summary is concluded in section IV.

\section{Brief description of the model}

In the LQMD model, the total wave function of the $N$-body nucleonic system is the direct product of each nucleon wave function by assuming a Gaussian wave packet, in which the fermionic nature of nucleons is neglected. After performing Wigner transformation, one can get the density distributions for nucleons in phase space, which are used to evaluate the Hamiltonian of the system based on the effective nucleon-nucleon interaction. The temporal evolutions of the baryons (nucleons and resonances) and mesons in the reaction system under the self-consistently generated mean-field are governed by Hamilton's equations of motion, which read as
\begin{eqnarray}
\dot{\mathbf{p}}_{i} = -\frac{\partial H}{\partial\mathbf{r}_{i}}, \quad \dot{\mathbf{r}}_{i} = \frac{\partial H}{\partial\mathbf{p}_{i}}.
\end{eqnarray}
The Hamiltonian of baryons consists of the relativistic energy, the Coulomb potential, the local interaction potential including the surface terms, and the momentum dependent interaction, which are evaluated with the Skyrme effective interactions \cite{Fe11}.

The density, isospin and momentum-dependent single-nucleon potential is obtained as follows \cite{Fe12}:
\begin{eqnarray}
U_{\tau}(\rho,\delta,\textbf{p}) = && \frac{1}{\rho_{0}}C_{\tau,\tau} \int d\textbf{p}' f_{\tau}(\textbf{r},\textbf{p})[\ln(\epsilon(\textbf{p}-\textbf{p}')^{2}+1)]^{2}         \nonumber \\
&&  + \frac{1}{\rho_{0}}C_{\tau,\tau'} \int d\textbf{p}' f_{\tau'}(\textbf{r},\textbf{p})[\ln(\epsilon(\textbf{p}-\textbf{p}')^{2}+1)]^{2}
\nonumber \\
&&  + \alpha\frac{\rho}{\rho_{0}}+\beta\frac{\rho^{\gamma}}{\rho_{0}^{\gamma}}+
E_{sym}^{loc}(\rho)\delta^{2} + \frac{\partial E_{sym}^{loc}(\rho)}{\partial\rho}\rho\delta^{2} \nonumber \\
&&  + E_{sym}^{loc}(\rho)\rho\frac{\partial\delta^{2}}{\partial\rho_{\tau}}  .
\end{eqnarray}
Here $\tau\neq\tau'$, $\partial\delta^{2}/\partial\rho_{n}=4\delta\rho_{p}/\rho^{2}$ and $\partial\delta^{2}/\partial\rho_{p}=-4\delta\rho_{n}/\rho^{2}$. The nucleon effective (Landau) mass in nuclear matter of isospin asymmetry $\delta=(\rho_{n}-\rho_{p})/(\rho_{n}+\rho_{p})$ with $\rho_{n}$ and $\rho_{p}$ being the neutron and proton density, respectively, is calculated through the potential as $m_{\tau}^{\ast}=m_{\tau}/ \left(1+\frac{m_{\tau}}{|\textbf{p}|}|\frac{dU_{\tau}}{d\textbf{p}}|\right)$ with the free mass $m_{\tau}$ at Fermi momentum $\textbf{p}=\textbf{p}_{F}$. The parameters $\alpha$, $\beta$, $\gamma$ and $\rho_{0}$ are set to be the values of -215.7 MeV, 142.4 MeV, 1.322 and 0.16 fm$^{-3}$, respectively. The $C_{\tau,\tau}=C_{mom}(1+x)$, $C_{\tau,\tau'}=C_{mom}(1-x)$ ($\tau\neq\tau'$) and the isospin symbols $\tau$($\tau'$) represent proton or neutron. The values of 1.76 MeV, 500 c$^{2}$/GeV$^{2}$ are taken for the $C_{mom}$ and $\epsilon$, respectively, which result in the effective mass $m^{\ast}/m$=0.75 in nuclear medium at saturation density for symmetric nuclear matter. The parameter $x$ as the strength of the isospin splitting with the value of -0.65 is taken in this work, which has the mass splitting of $m^{\ast}_{n}>m^{\ast}_{p}$ in nuclear medium.  A compression modulus of K=230 MeV for isospin symmetric nuclear matter is concluded in the LQMD model. 

The symmetry energy is composed of three parts, namely the kinetic energy coming from the difference of free Fermi-gas motion in the neutron-rich and isospin symmetric nuclear matters, the local density-dependent interaction and the momentum-dependent potential as
\begin{equation}
E_{sym}(\rho)=\frac{1}{3}\frac{\hbar^{2}}{2m}\left(\frac{3}{2}\pi^{2}\rho\right)^{2/3}+E_{sym}^{loc}(\rho)+E_{sym}^{mom}(\rho).
\end{equation}
The local part is adjusted to mimic predictions of the symmetry energy calculated by microscopical or phenomenological many-body theories and has two-type forms as follows:
\begin{equation}
E_{sym}^{loc}(\rho)=\frac{1}{2}C_{sym}(\rho/\rho_{0})^{\gamma_{s}},
\end{equation}
and
\begin{equation}
E_{sym}^{loc}(\rho)=a_{sym}(\rho/\rho_{0})+b_{sym}(\rho/\rho_{0})^{2}.
\end{equation}
The parameters $C_{sym}$, $a_{sym}$ and $b_{sym}$ are taken as the values of 52.5 MeV, 43 MeV, -16.75 MeV. The values of $\gamma_{s}$=0.5, 1, 2 lead to the soft, linear and hard symmetry energy in the domain of high densities, respectively, and the latter gives a supersoft symmetry energy at supra-saturation densities. All cases cross at saturation density with the value of 31.5 MeV. The density dependence of nuclear symmetry energy with the momentum-dependent interactions (MDI) is shown in Fig. 1. Different stiffness crosses at saturation density with the value of 31.5 MeV, which cover the largely uncertain of symmetry energy.

\begin{figure}
\includegraphics[width=8 cm]{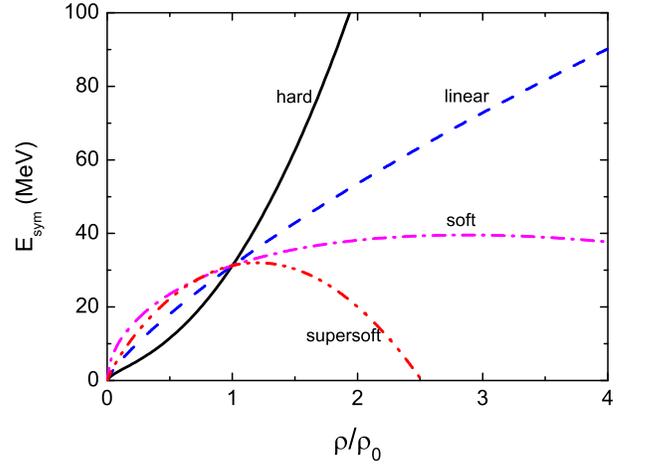}
\caption{\label{fig:epsart} (Color online) Density dependence of the nuclear symmetry energy for different stiffness with the MDI interaction.}
\end{figure}

The Hamiltonian of pions is constructed from the Coulomb potential and the energy of pion in medium as \cite{Fe15}
\begin{eqnarray}
H_{M} = \sum_{i=1}^{N_{M}}\left( V_{i}^{\textrm{Coul}} + \omega(\textbf{p}_{i},\rho_{i}) \right).
\end{eqnarray}
Here the Coulomb interaction is given by
\begin{equation}
V_{i}^{\textrm{Coul}}=\sum_{j=1}^{N_{B}}\frac{e_{i}e_{j}}{r_{ij}},
\end{equation}
where the $N_{M}$ and $N_{B}$ are the total numbers of mesons and baryons including charged resonances, respectively. The energy of pion in the nuclear medium is composed of the isoscalar and isovector contributions as follows
\begin{equation}
\omega_{\pi}(\textbf{p}_{i},\rho_{i}) = \omega_{isoscalar}(\textbf{p}_{i},\rho_{i})+C_{\pi}\tau_{z}\delta (\rho/\rho_{0})^{\gamma_{\pi}}.
\end{equation}
The coefficient $C_{\pi}= \rho_{0} \hbar^{3}/(4f^{2}_{\pi}) = 36$ MeV is taken from fitting the experimental data of pion-nucleus scattering \cite{Fr07}. The isospin quantities are taken as $\tau_{z}=$ 1, 0, and -1 for $\pi^{-}$, $\pi^{0}$ and $\pi^{+}$, respectively. The isospin asymmetry $\delta=(\rho_{n}-\rho_{p})/(\rho_{n}+\rho_{p})$ and the quantity $\gamma_{\pi}$ adjusts the stiffness of isospin splitting of the pion-nucleon potential. We take the $\gamma_{\pi} = 2$ in the work.

\begin{figure*}
\includegraphics[width=16 cm]{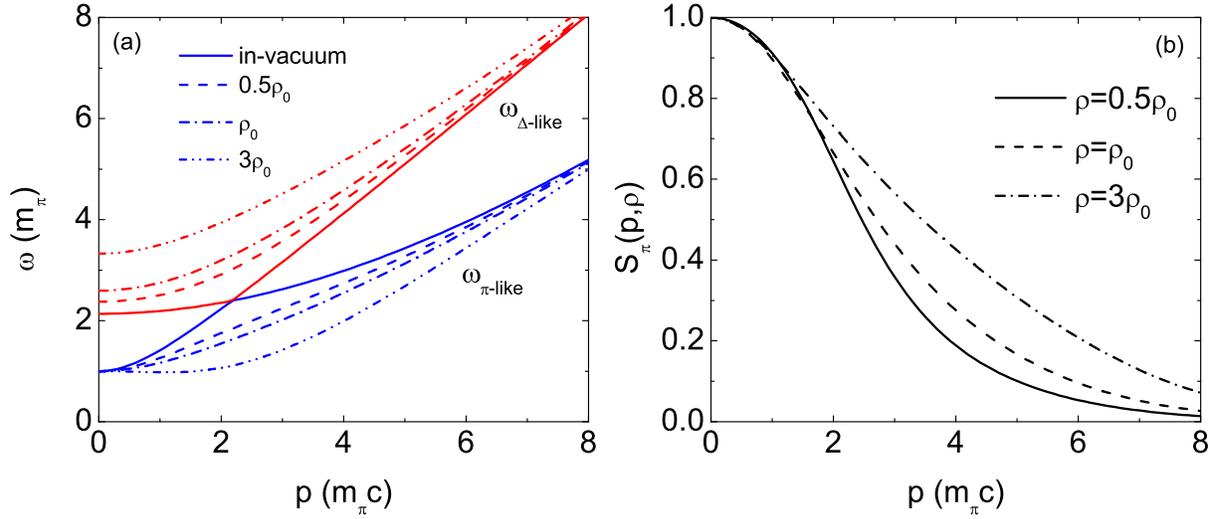}
\caption{\label{fig:wide} (Color online) The pion dispersion relation and the probabilities of the pion component in the pion branch as functions of pion momenta at different nuclear densities.}
\end{figure*}

\begin{figure*}
\includegraphics[width=16 cm]{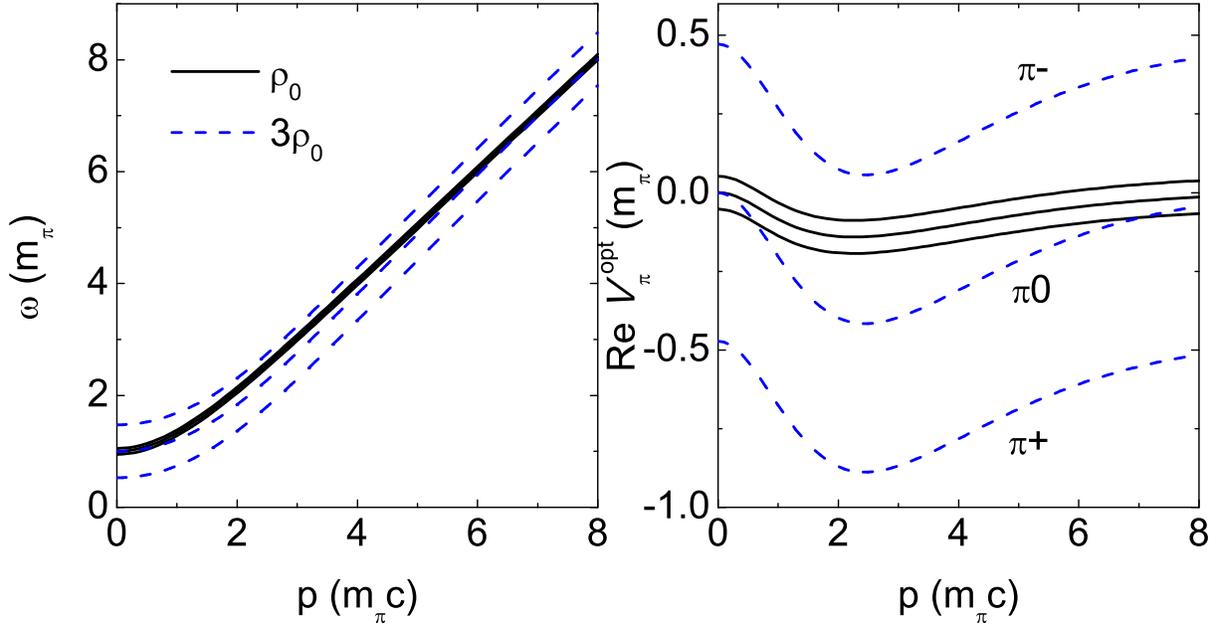}
\caption{\label{fig:wide} (Color online) Momentum dependence of the pion energy and the optical potential in dense nuclear matter.}
\end{figure*}

The pion self-energy in the nuclear medium have been studied via the $\Delta$-hole model \cite{Br75}. The in-medium pion dispersion relation consists of a pion branch (smaller value) and a $\Delta$-hole (larger value) branch, which become softened and hardened with baryon density in nuclear matter, respectively. Thus, the dispersion relation reads
\begin{eqnarray}
\omega_{isoscalar}(\textbf{p}_{i},\rho_{i}) = &&  S_{\pi}(\textbf{p}_{i},\rho_{i}) \omega_{\pi-like}(\textbf{p}_{i},\rho_{i}) +    \nonumber \\
&&        S_{\Delta}(\textbf{p}_{i},\rho_{i}) \omega_{\Delta-like}(\textbf{p}_{i},\rho_{i}) .
\end{eqnarray}
The probabilities of the pion component satisfy the relation
\begin{equation}
S_{\pi}(\textbf{p}_{i},\rho_{i}) + S_{\Delta}(\textbf{p}_{i},\rho_{i}) = 1
\end{equation}
The value of the probability is determined from the pion self-energy as \cite{Xi93}
\begin{equation}
S(\textbf{p}_{i},\rho_{i}) = \frac{1}{1-\partial \Pi (\omega)/\partial\omega^{2}},
\end{equation}
where the pion self-energy is given by
\begin{equation}
\Pi = \textbf{p}_{i}^{2}\frac{\chi}{1 - g\prime\chi},
\end{equation}
with the Migdal parameter $g\prime\sim$0.6 and
\begin{equation}
\chi = -\frac{8}{9}\left(\frac{f_{\Delta}}{m_{\pi}}\right)^{2} \frac{\omega_{\Delta}\rho\hbar^{3}}{\omega_{\Delta}^{2}-\omega^{2}} \exp(-2\textbf{p}_{i}^{2}/b^{2}).
\end{equation}
The $\omega_{\Delta}=\sqrt{m_{\Delta}^{2}+\textbf{p}_{i}^{2}}-m_{N}$, the $m_{\pi}$, $m_{N}$ and $m_{\Delta}$ are the masses of pion, nucleon and delta, respectively. The $\pi N\Delta$ coupling constant $f_{\Delta}\sim 2 $ and the cutoff factor $b\sim 7 m_{\pi}$. Two eigenvalues of $\omega_{\pi-like}$ and $\omega_{\Delta-like}$ are obtained from the pion dispersion relation as
\begin{equation}
\omega^{2} = \textbf{p}_{i}^{2} + m_{\pi}^{2} + \Pi(\omega).
\end{equation}
The $\pi-like$ and $\Delta-like$ in-medium energies and the pion-component probability at different baryon densities are shown in Fig. 2. The $\Delta-like$ energy increases with the baryon density and the $\pi-like$ energy exhibits an opposite trend. The both cases cross at the $\Delta$ resonance energy, i.e., $E_{lab}=$0.19 GeV or $p=$0.298 GeV/c, which lead to a pocket appearance in the pion optical varying with the pion momentum. The pion-component contribution to the energy in nuclear medium is reduced with increasing the pion momentum. The energy balance in the decay of resonance is satisfied with the relation
\begin{equation}
\sqrt{m_{R}^{2}+\textbf{p}_{R}^{2}} = \sqrt{m_{N}^{2}+(\textbf{p}_{R} -\textbf{p}_{\pi})^{2}} + \omega_{\pi}(\textbf{p}_{\pi},\rho),
\end{equation}
where the $\textbf{p}_{R}$ and $\textbf{p}_{\pi}$ are the momenta of resonances and pions, respectively.
The optical potential can be evaluated from the in-medium energy $V_{\pi}^{opt}(\textbf{p}_{i},\rho_{i}) = \omega_{\pi}(\textbf{p}_{i},\rho_{i}) - \sqrt{m_{\pi}^{2}+\textbf{p}_{i}^{2}}$. Shown in Fig. 3 is the pion energy and optical potential as a functions of the pion momentum at the baryon densities of $\rho_{0}$ and 3$\rho_{0}$, respectively. The strengths of the potentials at the $\Delta$ resonance energy ($p=2.1m_{\pi}$) are -12 MeV, -19 MeV and -27 MeV at the normal density, but 7 MeV, -58 MeV and -124 MeV at the baryon density of $3\rho_{0}$ for $\pi^{-}$, $\pi^{0}$ and $\pi^{+}$, respectively. Isospin splitting of the pion potential is pronounced at high baryon densities, which will impact the pion dynamics in heavy-ion collisions, in particular the $\pi^{-}/\pi^{+}$ ratio.

The probability in two-particle collisions to a channel is performed by using a Monte Carlo procedure by its contribution of the channel cross section to the total cross section. I have included the reaction channels for the pion production as follows:
\begin{eqnarray}
&& NN \leftrightarrow N\triangle, \quad  NN \leftrightarrow NN^{\ast}, \quad  NN
\leftrightarrow \triangle\triangle,  \nonumber \\
&& \Delta \leftrightarrow N\pi,  N^{\ast} \leftrightarrow N\pi,  NN \leftrightarrow NN\pi (s-state).
\end{eqnarray}
The momentum-dependent decay widths are used for the resonances of $\Delta$(1232) and $N^{\ast}$(1440) \cite{Hu94}. We have taken a constant width of $\Gamma$=150 MeV for the $N^{\ast}$(1535) decay. Elastic scatterings in nucleon-nucleon, nucleon-resonance ($NR\rightarrow NR$) and resonance-resonance ($RR\rightarrow RR$) collisions and inelastic collisions of nucleon-resonance ($NR\rightarrow NN$, $NR\rightarrow NR\prime$) and resonance-resonance ($RR\rightarrow NN$, $RR\rightarrow RR\prime$, $R$ and $R\prime$ being different resonances), have been included in the model.

The cross section of pion-nucleon scattering is evaluated with the Breit-Wigner formula as the form of
\begin{eqnarray}
\sigma_{\pi N\rightarrow R}(\sqrt{s}) = \sigma_{max}(|\textbf{p}_{0}/\textbf{p}|)^{2}\frac{0.25\Gamma^{2}(\textbf{p})}
{0.25\Gamma^{2}(\textbf{p})+(\sqrt{s}-m_{0})^{2}},
\end{eqnarray}
where the $\textbf{p}$ and $\textbf{p}_{0}$ are the momenta of pions at the energies of $\sqrt{s}$ and $m_{0}$, respectively, and $m_{0}$ being the centroid of resonance mass, e.g., 1.232 GeV, 1.44 GeV and 1.535 GeV for $\Delta$(1232), $N^{\ast}$(1440), and $N^{\ast}$(1535), respectively. The maximum cross section $\sigma_{max}$ is taken from fitting the cross sections of the available experimental data in pion-nucleon scattering \cite{Ol14}.

\section{Results and discussion}

\begin{figure}
\includegraphics[width=8 cm]{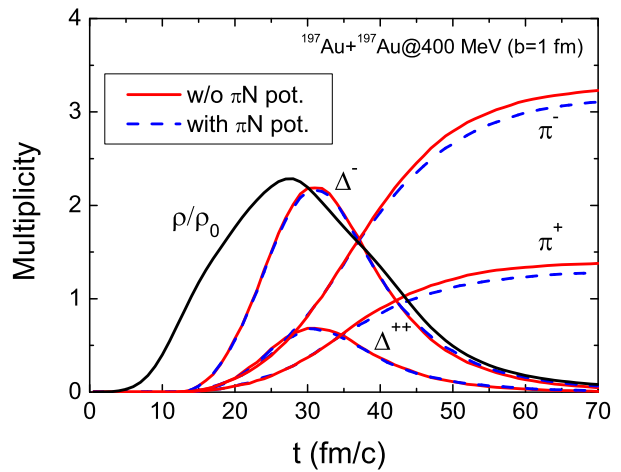}
\caption{\label{fig:epsart} (Color online)  Temporal evolutions of central density, charged pions and $\Delta(1232)$ in central $^{197}$Au+$^{197}$Au collisions at an incident energy of 400\emph{A} MeV.}
\end{figure}

\begin{figure*}
\includegraphics[width=16 cm]{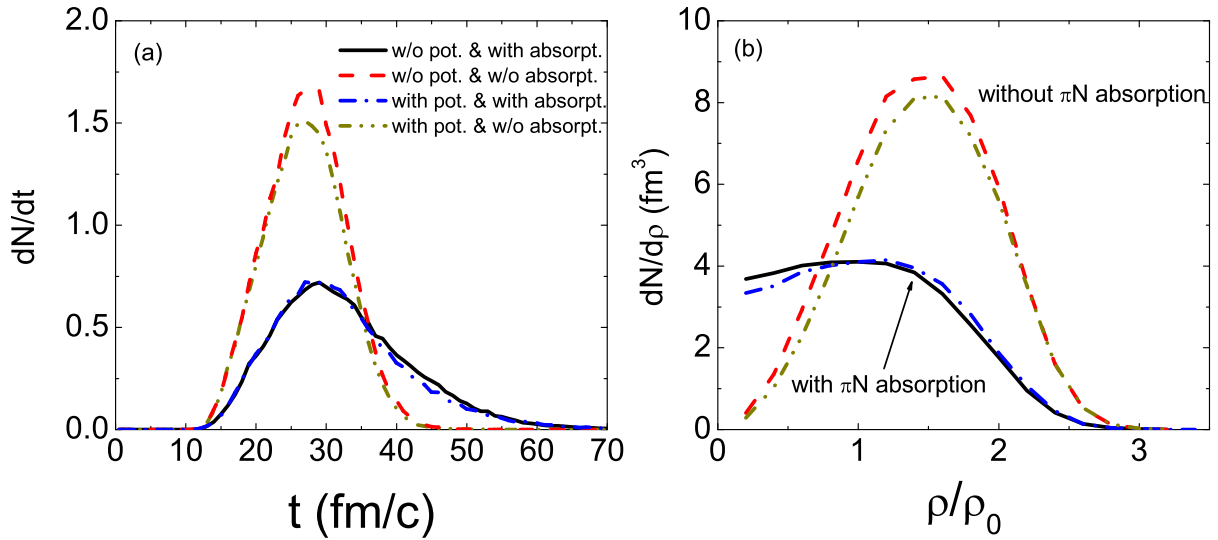}\caption{\label{fig:wide} (Color online) Impacts of the reabsorption channels in pion-nucleon collisions and of pion-nucleon potential on the production rate and density profiles in central $^{197}$Au+$^{197}$Au collisions at the incident energy of 400\emph{A} MeV.}
\end{figure*}

Pion as the lightest meson in the nature can be produced in nucleon-nucleon collisions with a lower threshold energy (E$_{th}$=289 MeV for $\pi^{\pm}$). The stochastic processes in heavy-ion collisions accelerate part of nucleons, which make the particle production at sub-threshold energies. Dynamics of pions produced in heavy-ion collisions has been investigated both in theories and in experiments, which is related to the issues of the high-density symmetry energy and the pion-nucleon interaction in dense matter. Shown in Fig. 4 is the impacts of pion-nucleon potential on the production of charged pions and $\Delta(1232)$ in central $^{197}$Au+$^{197}$Au collisions at an incident energy of 400\emph{A} MeV. It is should to be noticed that both multiplicities of $\pi^{-}$ and $\pi^{+}$ are slightly reduced with the pion potential. The absorption collisions of pions and nucleons via the channel of $\pi N\rightarrow \Delta(1232)$ dominate the pion dynamics in heavy-ion collisions. The production rate and density profile of pions produced in central $^{197}$Au+$^{197}$Au collisions at the incident energy of 400\emph{A} MeV is shown in Fig. 5. The reabsorption process retards the pion production towards the subnormal densities. The maximal yields correspond to the densities of 1.0$\rho_{0}$ and 1.6$\rho_{0}$ with and without the pion-nucleon ($\pi$N) scattering, respectively. The average density in the production pion is evaluated by $<\rho>=\int\rho dN/N_{total}$ and $N_{total}$ being the number of total pions, and to be 1.07$\rho_{0}$ and 1.44$\rho_{0}$ with and without the $\pi$N absorption, respectively. Therefore, pions produced in heavy-ion collisions provide the information of phase diagram around saturation densities.

\begin{figure*}
\includegraphics[width=16 cm]{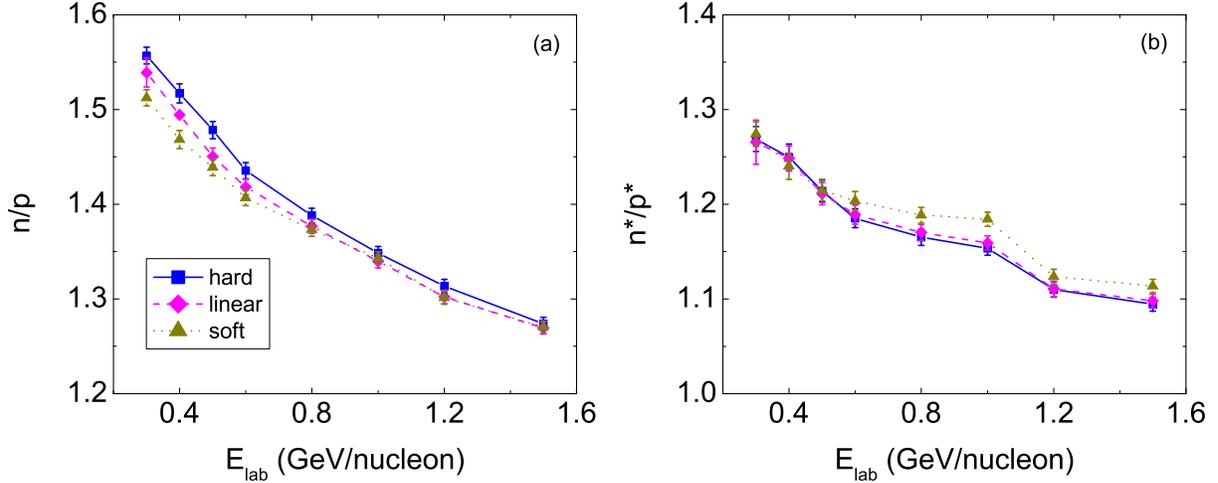}
\caption{\label{fig:wide} (Color online) The $n/p$ and $n^{\ast}/p^{\ast}$ ratios from free the preequilibrium light particles within the rapidity bin of $|y/y_{proj}|<$0.2 and within the azimuthal angles of $70^{o} \sim 110^{o}$ and $250^{o} \sim 290^{o}$ in the $^{197}$Au+$^{197}$Au reaction at the energy of 300 MeV/nucleon.}
\end{figure*}

Preequilibrium nucleons in the domain of squeezed out in heavy-ion collisions could bring the high-density information of nuclear matter. The fast nucleons is perpendicular to the reaction plane and counted within the rapidity bin of $|y/y_{proj}|<$0.2 and within the azimuthal angles of $70^{o} \sim 110^{o}$ and $250^{o} \sim 290^{o}$. The $n/p$ and $n^{\ast}/p^{\ast}$ ratios of free nucleons and ¡¯gas-phase¡¯ nucleons
(nucleons, hydrogen and helium isotopes) from the preequilibrium particles is shown in Fig. 6. The isospin effects of the $n/p$ ratios are pronounced at the energies below 0.8$A$ GeV. The value increases with the stiffness factor $\gamma_{s}$ because of more repulsive for neutrons with hardening the symmetry energy at the supra-saturation densities.

\begin{figure*}
\includegraphics[width=16 cm]{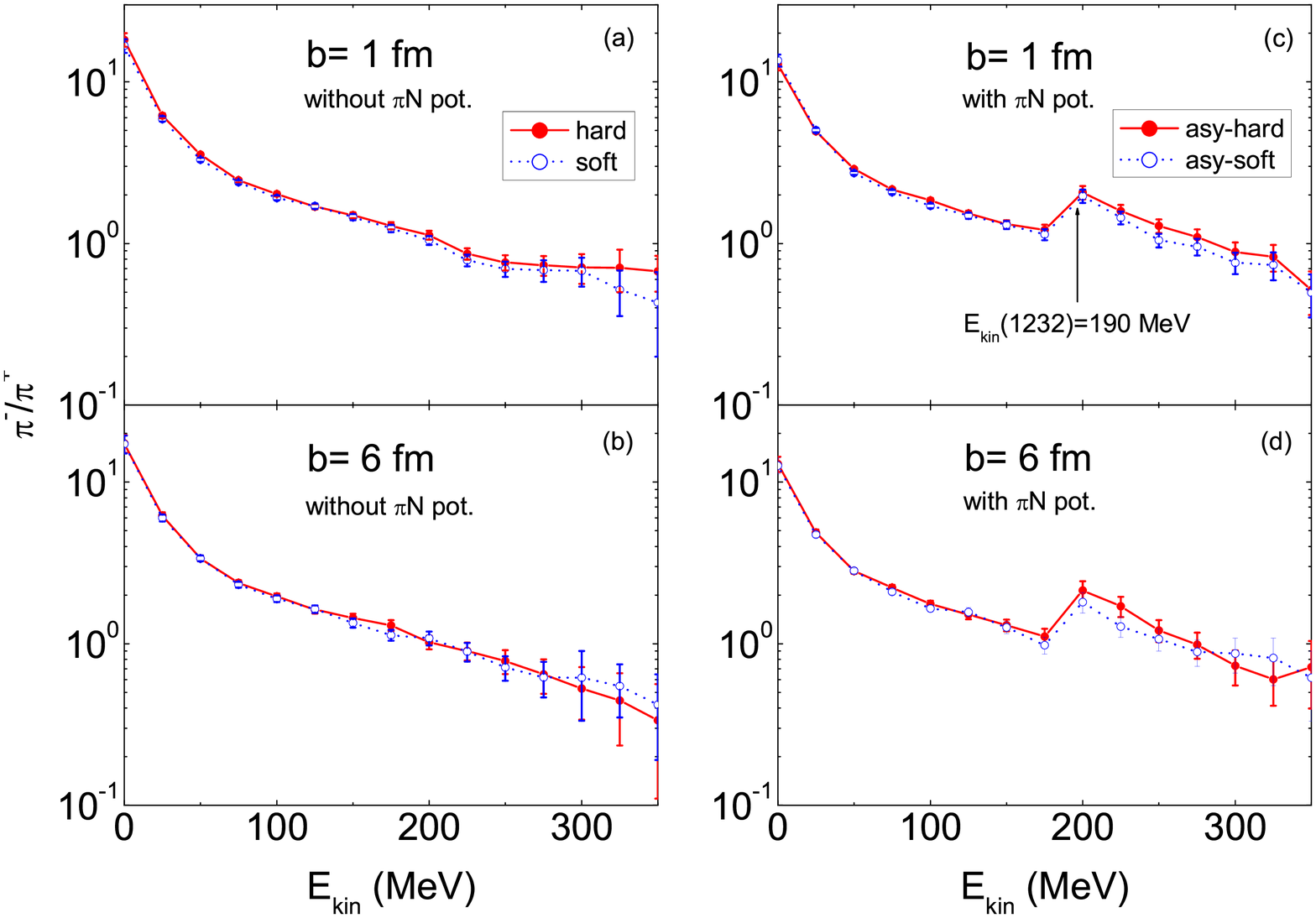}
\caption{\label{fig:wide} Impacts of collision centrality, stiffness of symmetry energy and pion-nucleon potential on kinetic energy spectra of the $\pi^{-}/\pi^{+}$ ratios in the $^{197}$Au+$^{197}$Au reaction at an incident energy of 300 MeV/nucleon.}
\end{figure*}

\begin{figure*}
\includegraphics[width=16 cm]{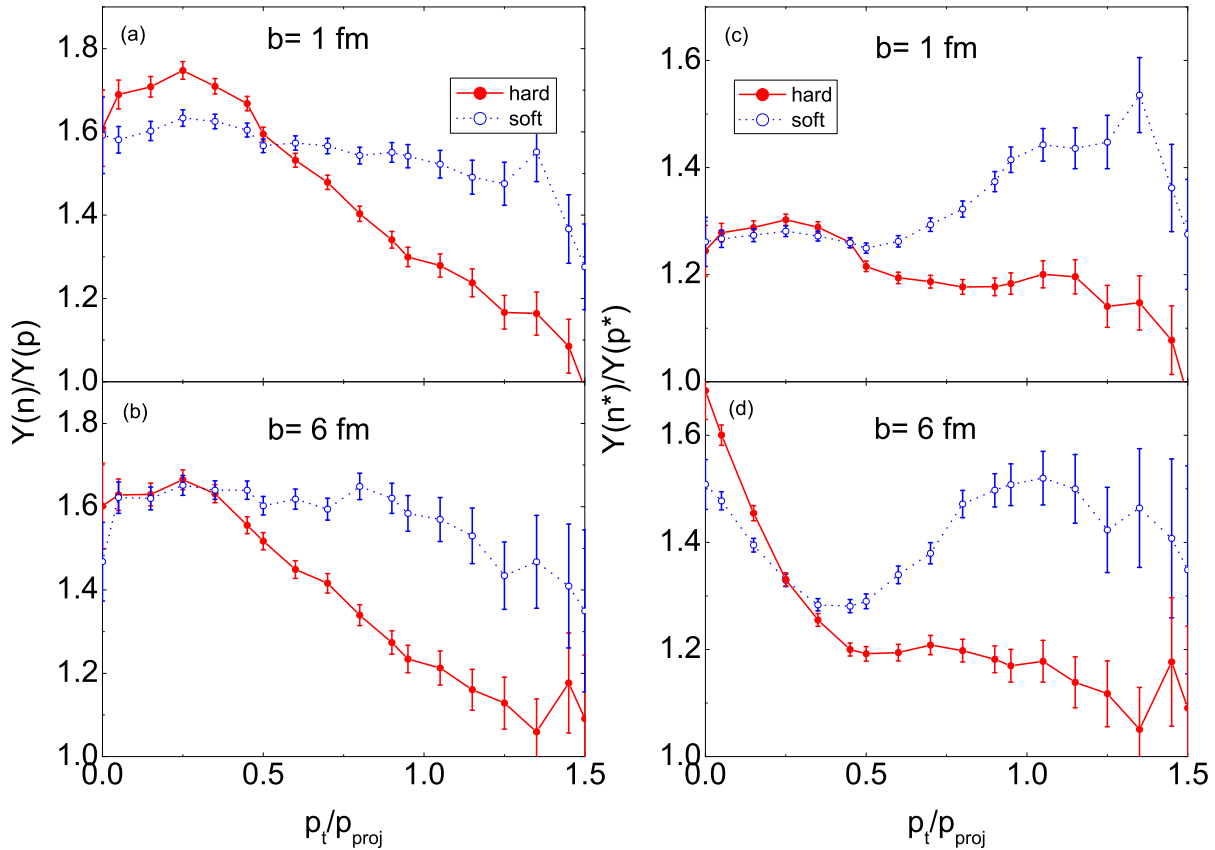}
\caption{\label{fig:wide} Transverse mass spectra of the $n/p$ and $n^{\ast}/p^{\ast}$ ratios in the $^{197}$Au+$^{197}$Au reaction at the energy of 300 MeV/nucleon.}
\end{figure*}

\begin{figure*}
\includegraphics[width=16 cm]{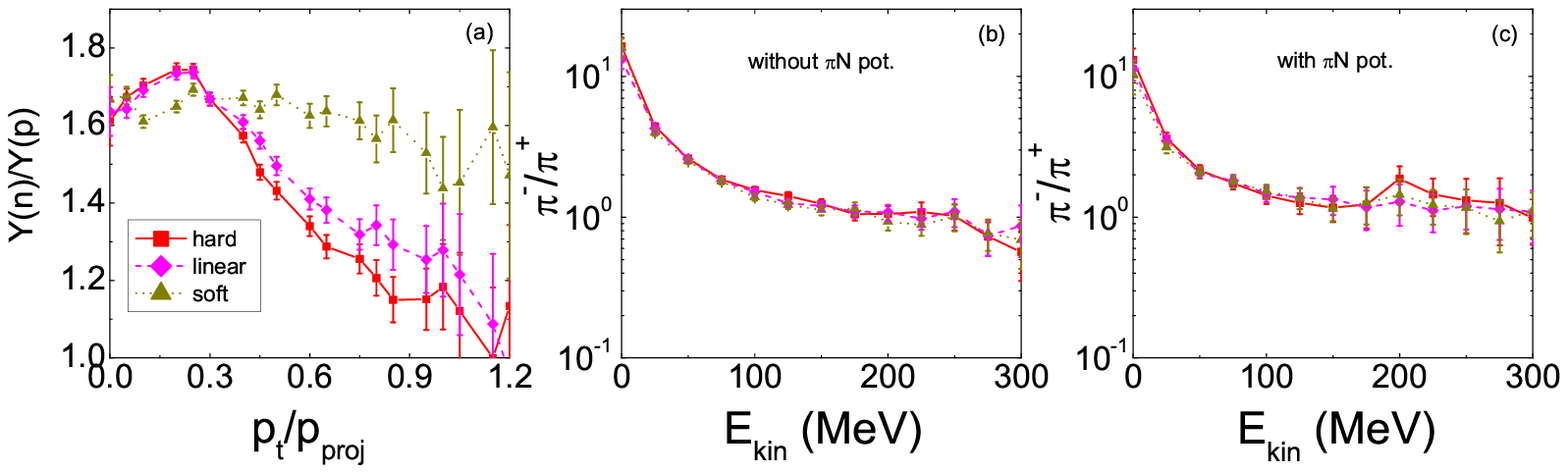}
\caption{\label{fig:wide} The $n/p$ and $\pi^{-}/\pi^{+}$ ratios in central $^{40}$Ar+$^{197}$Au collisions at an incident energy of 300 MeV/nucleon with different stiffness of symmetry energies as functions of transverse momentum and kinetic energy, respectively.}
\end{figure*}

Kinetic energy spectra of particles in heavy-ion collisions will bring the hidden information of production mechanism in comparison to the total yields, i.e., the mean-field potential, reaction channels, elementary cross sections etc. Shown in Fig. 7 is the kinetic energy spectra of the $\pi^{-}/\pi^{+}$ ratios in the $^{197}$Au+$^{197}$Au reaction at the incident energy of 300 MeV/nucleon. It is interest to be noticed that a peak structure appears at the $\Delta(1232)$ resonance energy ($E_{kin}$=190 MeV) with inclusion of the pion-nucleon potential. The conclusions depend on the formulation of the pion-nucleon interaction. However, the isospin effect is negligible in the kinetic energy spectra. Calculations from the Boltzmann-Uehling-Uhlenbeck (pBUU) transport model presented an eyeable effect of the $\pi^{-}/\pi^{+}$ ratio varying with the stiffness of symmetry energy at high kinetic energies ($>$80 MeV) \cite{Ho14}. The kinetic energy spectra of the $\pi^{-}/\pi^{+}$ ratios are expected in the near future experiments such as HIRFL-CSR (Lanzhou), RIKEN-SAMURAI in Japan etc, not only for the high-density symmetry energy, also for the pion-nucleon interaction in dense nuclear matter. Shown in Fig. 8 is the transverse mass spectra of the $n/p$ and $n^{\ast}/p^{\ast}$ ratios from the preequilibrium particles in the $^{197}$Au+$^{197}$Au reaction at the energy of 300 MeV/nucleon. The soft symmetry energy has a flat structure. The spectra are different with the neck fragmentation in fermi-energy heavy-ion collisions \cite{Fe16}, which is used to extract the subsaturation-density symmetry energy. The stiffness of symmetry energies on the $n/p$ and $\pi^{-}/\pi^{+}$ ratios in central $^{40}$Ar+$^{197}$Au is shown Fig. 9. Very similar structure is observed with $^{197}$Au+$^{197}$Au collisions. The system has particularly interests in the future HIRFL-CSR (Lanzhou) experiments.

\section{Conclusions}

The dynamics of pions and preequilibrium particles produced in heavy-ion collisions to probe the high-density symmetry energy, has been investigated within an isospin and momentum dependent transport model (LQMD). The reabsorption process in pion-nucleon collisions retards the pion production towards the saturation-density region. The isospin dependent pion-nucleon potential slightly reduces the pion yields, but enhances the $\pi^{-}/\pi^{+}$ ratio, in particular in the domain of subthreshold energies. A bump structure appears in the kinetic energy spectra at the $\Delta(1232)$ resonance energy with the potential. The neutron to proton ratios of the preequilibrium particles could be nice probes for extracting the high-density symmetry energy in heavy-ion collisions, i.e., transverse momentum (kinetic energy) spectra, yield ratios such as $n/p$ and $n^{\ast}/p^{\ast}$.

\section{Acknowledgements}

This work was supported by the Major State Basic Research Development Program in China (No. 2014CB845405 and No. 2015CB856903), and the National Natural Science Foundation of China (Projects No. 11675226, No. 11175218, and No. U1332207).

\end{document}